%% file: main.tex
\documentclass[aps,onecolumn,showpacs,superscriptaddress,prl,11pt]{revtex4-1}

\input{packages.tex}
\input{macros.tex}
\input{specifications.tex}

\begin{document}

\title{Dynamical BCS theory of a two-dimensional attractive Fermi gas: effective interactions from Quantum Monte Carlo calculations}

\author{Ettore Vitali}
\affiliation{Department of Physics, California State University Fresno, Fresno, California 93740}

\author{Jimmy Gonzalez}
\affiliation{Department of Physics, California State University Fresno, Fresno, California 93740}

\keywords{Ultracold, Fermi, Atomic, Strongly-Correlated, QMC, BCS}
\begin{abstract}
    \input{abstract.tex}
\end{abstract}

\maketitle
\section{Introduction}
    \input{introduction.tex}

\section{Quick review of the Dynamical BCS Theory}
    \input{dbcs_theory.tex}
    \input{dbcs_implementation.tex}

\section{Comparison with exact results and renormalization of the effective interaction}
    \input{calculations.tex}

\section{Discussion and Conclusion}
    \input{conclusion.tex}
\input{main.bbl}
\end{document}

%% file: packages.tex
\usepackage[titlenumbered,ruled]{algorithm2e}
\usepackage{algpseudocode}
\usepackage{graphicx}
\usepackage{dcolumn}
\usepackage{bm}
\usepackage{color}
\usepackage{amsmath}
\usepackage{amsfonts}
\usepackage{subfig}
\usepackage[makeroom]{cancel}
\usepackage{float}
\usepackage{letltxmacro}
\usepackage{caption}

\usepackage{physics} 

%% file: macros.tex
\newcommand{\COMMENTED}[1]{}

%% file: specifications.tex
\everymath{\displaystyle} 

\graphicspath{Images/}

%% file: abstract.tex
The primary work presented in this paper focuses on the calculation of density-density dynamical correlations in an attractive two dimensional Fermi gas 
in several physically interesting regimes, including the strongly correlated BEC-BCS crossover regime. 
We use state-of-the-art dynamical BCS theory and we address the possibility to renormalize the interaction strength, using unbiased Quantum Monte Carlo results as an asset to validate the predictions. We propose that a suitable interplay between dynamical BCS theory, which is computationally very cheap and yields results directly in real time domain, and Quantum Monte Carlo methods, which are exact but way more demanding and limited to imaginary time domain, can be a very promising idea to study dynamics in many body systems. We illustrate the idea and provide quantitative results for a few values of the interaction strength in the cold gas.


%% file: introduction.tex
Fermionic superfluids, both in three dimensions and in lower dimensionality, have become a very important focus of research 
, with cold atomic Fermi systems providing a very exciting opportunity to observe quantum mechanics at work. 
Thanks to remarkable improvements in the technology, it is now possible to produce atomic gas systems at very low temperatures, and to visualize quantum mechanics with an impressive resolution and control (we refer the readers, for example, to the review papers \cite{RevModPhys.80.1215,RevModPhys.80.885}). The study of those unique gases, where it is possible to tune the interatomic forces through Feshbach resonances \cite{PhysRevA.78.053606},  provides a deep insight into quantum phenomena, with applications in unconventional superconductivity, nuclear physics, and nuclear astrophysics  \cite{RevModPhys.78.17,RevModPhys.83.1057, PhysRevLett.90.161101, PhysRevLett.90.161101}.
Another key parameter that can be tuned when we study those systems is the dimensionality:
using anisotropic potentials \cite{PhysRevLett.105.030404}, it is now possible to confine the gases
in two dimension (2D). The effect is highly not trivial: the existence of bound states for pairs of fermions 
for arbitrary attractive potentials \cite{Galea2017}, which is  
a unique feature of 2D gases, together with the expected enhancement of quantum fluctuations
make the study of cold atoms moving in 2D very interesting. 

In this paper we focus on an ultracold attractive Fermi atomic gas, for example made of $^6$Li atoms, confined to move in 2D. We address the delicate problem of the theoretical calculation of the density response function of the system. Obtaining response functions for quantum systems from first principles is a major challenge, the importance being due to the fact that those quantities can be directly compared with experiments, and in particular with scattering experiments \cite{PhysRevLett.109.050403}. Focusing on the dilute limit, which is the typical condition for cold atoms, we will study the density response function in various regimes, namely the BCS, BEC, and the BEC-BCS crossover, the latter being an exciting example of a strongly correlated system \cite{RevModPhys.80.885}.
These regimes are of special interest as the physical properties dramatically change as a function of the interaction strength: in a BCS superfluid the interaction  is weak, and the system resembles an ordinary superconductor with the instability of the Fermi sea lending toward the formation of Cooper pairs close to the Fermi surface; as the interaction becomes stronger, the system continuously changes to a BEC superfluid, where the interaction between atoms is sufficiently strong that pairs become tightly bound, forming Bosonic molecules. Between these two regimes lies the strongly correlated BEC-BCS crossover, in which, in some sense, the superfluid behaves as a quantum superposition of weakly bound Cooper pairs and tightly bound molecules. 
A very interesting question is how does the density response function change, as we move across such a rich phase diagram. 

It is well known that, even when we address the study of static properties, it is very hard for simple theoretical approaches to provide robust quantitative results, in particular in the crossover regime where the correlations are strong and simple perturbative approaches are doomed to fail. Quantum Monte Carlo simulations, on the other hand, and in particular Auxiliary Field Quantum Monte Carlo (AFQMC), are able to provide exact numerical results for physical properties in 2D, such as the equation of state, the momentum distribution, the pairing wave function \cite{PhysRevA.92.033603} and, with more recent developments, also pairing gaps \cite{PhysRevA.96.061601}. Thanks to the great success of AFQMC, one might argue that the problem of a 2D attractive Fermi gas is solved, and there is no need for additional studies. Although this might be true for the static properties at zero temperature, the dynamical properties are still a huge challenge. In \cite{PhysRevA.96.061601} it was shown that AFQMC can provide exact estimations also for dynamics, which is a major advance, but two important observations need to be considered: first of all, AFQMC allows us to compute exact dynamical correlations only in imaginary time domain, and the mapping to real time domain is far from being trivial (see for example \cite{statisticalcomp}). Moreover, although the scaling of the AFQMC technique \cite{PhysRevB.94.085140} is very favorable, it remains true that the extrapolation to the infinite system limit is very demanding.
It would certainly be very desirable to have a simple theory that allows us to compute a reasonable approximation for the density response function, both providing results directly in real time domain and allowing us to compute properties in the thermodynamic limit with minimal computational effort. 
Normally this is achieved by mean field approaches and the difficulty lies in the fact that, in general, the underlying approximations are uncontrollable and it is difficult to draw robust physical conclusions. 

This is exactly the motivation for our study: in this work we will build on our previous study \cite{responsefunctions}, where we made a simple comparison between AFQMC and the state-of-art dynamical BCS theory \cite{PhysRevA.74.042717}. We explore whether it is possible to use the exact AFQMC results in imaginary time as an asset to assess and improve the dynamical BCS predictions through a suitable renormalization of the parameters.
The advent of those novel methodologies, like AFQMC, which are sometimes able to give us exact answers, makes it crucial and timely, in our opinion, to investigate the idea of interfacing those techniques with simple methodologies, in order to maximize the benefits, like having results in real time domain and being able to better control the accuracy, while minimizing the side effects, like the need of analytical continuation, the computational effort or the uncontrolled approximations. Our work is just a first simple step in this direction, but we hope that it could motivate further reasearch.

In the following we will review the key elements of the dynamical BCS theory, in order to help a reader implement the dynamical BCS methodology as a powerful tool to study response functions in cold atomic systems.
Through a detailed comparison with exact AFQMC predictions, we investigate the possibility of obtaining the density response function through the dynamical BCS theory using physical and renormalized effective parameters. 

The rest paper is organized as follows: we first review and highlight key points in the development of the dynamical BCS theory; then, we discuss how dynamical BCS is implemented. In the following section we examine the possibility to use effective parameters and the improvement these provide when compared with exact AFQMC results. Finally, we draw our conclusions.

%% file: dbcs_theory.tex
In this section we will set our notations, we will describe a few key points of dynamical BCS theory, and we will describe the essential ingredients that are needed for an implementation of the approach.
We use the customary spin-up and spin-down labels, $\ket{\uparrow}, \ket{\downarrow}$, to refer to two hyperfine states, for example the two lowest energy hyperfine states of $^6$Li. The particles are modeled as structureless $1/2$-spin fermions and atomic interactions are simple two-body zero-range s-wave interactions, for example  $V _ { \uparrow \downarrow } \left( \mathbf { r } , \mathbf { r } ^ { \prime } \right) = g \delta \left( \mathbf { r } - \mathbf { r } ^ { \prime } \right)$; the Hamiltonian for cold atoms is thus written as
    \begin{align} \label{eqn: cold atoms}
        \begin{aligned} H := & \int d \mathbf { r } \sum _ { \sigma = \uparrow , \downarrow } \psi _ { \sigma } ^ { \dagger } ( \mathbf { r } ) \left( - \frac { \hbar ^ { 2 } \nabla ^ { 2 } } { 2 m } - \mu \right) \psi _ { \sigma } ( \mathbf { r } ) + \\ & + g \int d \mathbf { r } \psi _ { \uparrow } ^ { \dagger } ( \mathbf { r } ) \psi _ { \downarrow } ^ { \dagger } ( \mathbf { r } ) \psi _ { \downarrow } ( \mathbf { r } ) \psi _ { \uparrow } ( \mathbf { r } ) \end{aligned}.
    \end{align}
    where the parameter $g$ measures the interaction strength.
    
    It turns out to be useful to recast the above Hamiltonian in momentum space by expanding the field operators in plane waves via
    \begin{equation}
        \psi _ { \sigma } ( \mathbf { r } ) = \frac { 1 } { \sqrt { V } } \sum _ { \mathbf { k } } e ^ { i \mathbf { k } \mathbf { r } } c _ { \mathbf { k } \sigma }.
    \end{equation}
    Here a super-cell $\Omega = [ - L / 2 , L / 2 ] \times [ - L / 2 , L / 2 ] $ of volume $V = L^2$ has been introduced and periodic boundary conditions are used, giving rise to discrete momenta $\mathbf{k} = \frac{2\pi}{L} \mathbf{n}$, where $\mathbf{n} \in \mathbb{Z}^2$ is a vector of integers.
    The Hamiltonian becomes \begin{equation} \label{eqn:regularized hubbard hamiltonian}
        \begin{aligned} \hat { H } = \sum _ { \mathbf { k } , \sigma = \uparrow , \downarrow } \left( \frac { \hbar ^ { 2 } | \mathbf { k } | ^ { 2 } } { 2 m } - \mu \right) \hat { c } _ { \mathbf { k } , \sigma } ^ { \dagger } \hat { c } _ { \mathbf { k } , \sigma }  - \frac { g } { V } \sum _ { \mathbf { k } , \mathbf { k } ^ { \prime } , \lambda } \hat { c } _ { \mathbf { k } + \lambda , \uparrow } ^ { \dagger } \hat { c } _ { - \mathbf { k } + \lambda , \downarrow } ^ { \dagger } \hat { c } _ { - \mathbf { k } ^ { \prime } , \downarrow } \hat { c } _ { \mathbf { k } ^ { \prime } . \uparrow } \end{aligned}
    \end{equation}
    
    Due to unphysical divergences that arise from the zero-range interaction, it is necessary to introduce some additional regularization strategy. We use a lattice regularization, restricting the summations over the momenta to the first Brillouin zone $[-\pi/b, \pi/b] \times [-\pi/b, \pi/b]$ of a square lattice of parameter $b$. This is equivalent to constraining the system to move on the above mentioned lattice. The original Hamiltonian can be recovered by performing the continuum limit $b \to 0$ together with the thermodynamic limit $L \to +\infty$ \cite{responsefunctions}.
     The \emph{regularized} Hamiltonian in momentum space is 
     the Hubbard Hamiltonian for cold atoms, with a modified dispersion relation proportional to $|\mathbf{k}|^2$, consistent with the continuum limit and the diluteness condition, the average number of particles being always much smaller than the number of lattice sites.
     When we regularize the system using an auxiliary lattice, as we do here, the interaction parameter needs to be suitably modified to guarantee the correct continuum and thermodynamic limit. If $a$ is the scattering length \cite{Galea2017} of the interaction potential and $n$ is the density of the gas, which is customary expressed in terms of the Fermi momentum defined by $k_F = \sqrt{2\pi n}$, in the expression \ref{eqn:regularized hubbard hamiltonian} we will need to replace \cite{PhysRevA.86.013626}
     \begin{equation} \label{eqn:HubbardU}
         g \rightarrow  V_0  = t \, \frac { 4 \pi } { \log \left( k _ { F } a \right) - \log ( 0.80261 \sqrt { n } ) }, \quad t = \frac { \hbar ^ { 2 }  } { 2 m } .
     \end{equation}
     The dimensionless quantity $\log \left( k _ { F } a \right)$, in 2D, is the natural quantity to describe the interaction strength, as we will discuss in more detail below.
     
     

    The well known BCS theory starts from a mean-field break-up, which replaces operators with operator averages in the Hamiltonian \ref{eqn:regularized hubbard hamiltonian}), and yields the approximate interaction potential, $\hat{V}_{BCS}$, that describes Cooper pairs of zero center of mass momentum,
    \begin{equation} \label{eqn: BCS potential}
        \hat { V } \simeq \hat { V } _ { B C S } = - \frac { V | \Delta | ^ { 2 } } { V_0 } + \sum _ { \mathbf { k } } \Delta  \hat { c } _ { - \mathbf { k } , \downarrow } \hat { c } _ { \mathbf { k } , \uparrow } + \Delta \hat { c } _ { \mathbf { k } , \uparrow } ^ { \dagger } \hat { c } _ { \mathbf { - k } , \downarrow } ^ { \dagger }.
    \end{equation}
    The quantity $\Delta$, which is taken to be real, that appears in equation \ref{eqn: BCS potential}, is called the pairing gap and satisfies the following gap equation
    \begin{equation}
        1 = -\frac{V_0}{V}\sum_k \frac{1}{2E_k},
    \end{equation}
    where the quasi-particles energies are $E_k := \sqrt{(e_k -\mu)^2 + \Delta^2 }$, the free particle dispersion being given by $e_k = \frac { \hbar ^ { 2 } | \mathbf { k } | ^ { 2 } } { 2 m } $.
    The constant $\mu$, the chemical potential, is not the same as in \ref{eqn:regularized hubbard hamiltonian}, but it is determined self-consistently by imposing that the average particle density computed on the ground state of the mean-field hamiltonian
   corresponds to the desired density of the gas we are studying. This leads to a particle number equation which, coupled with the gap equation, allows us to compute the chemical potential and the pairing gap for a given strength of the interaction. Having in mind a practical implementation, the first important step is to solve the gap and particle number equations to find $(\Delta,\mu)$ for the given $V_0$, or $\log(k_F a)$, and $n=N/L^2$, the latter being the desired particle density. Here $N$ denotes the average number of particles.

    When we turn to dynamical properties, we embed the system in an external time dependent potential
    $\hat{U}( t )$. The key idea of the dynamical BCS theory is that, at any time $t$, the system will generate, self-consistently, a dynamical pairing field, as the Cooper pair will respond to the external perturbation.
    This requires a mean-field breakup which is more sophisticated than the simple one presented above.
    The dynamical BCS theory implements this idea in the framework of linear response theory, assuming that the perturbation is weak enough so that we can linearize the self-consistent equations.
    In practice, the perturbation, coupled with the particle density, is written in the form
    \begin{equation}
        \hat{U}(t)=\frac{1}{V} \delta U(\mathbf{q}, \omega) e^{-i \omega t} \sum_{\mathbf{k}} \sum_{\sigma} \hat{c}_{\mathbf{k}-\mathbf{q} / 2, \sigma}^{\dagger} \hat{c}_{\mathbf{k}+\mathbf{q} / 2, \sigma}+h.c.,
    \end{equation}
    and we study a dynamical mean-field Hamiltonian $\hat{H}_{BCS}(t)$ which contains a 
    time-dependent gap, 
    \begin{equation} \label{eqn: dynamical gap}
        \Delta(\mathbf{q},t)=-\frac{g}{V} \sum_{\mathbf{k}}\left\langle\hat{c}_{-(\mathbf{k}-\mathbf{q} / 2), \downarrow} \hat{c}_{\mathbf{k}+\mathbf{q} / 2 ), \uparrow}\right\rangle.
    \end{equation}
    We observe from \eqref{eqn: dynamical gap} that Cooper pairs with total momentum $\mathbf{q}$ can be formed, due to the presence of the perturbation \cite{PhysRevA.74.042717}.
    The time-dependent gap, which satisfies a time-dependent gap equation, and the static gap, playing the role of our $0^{th}$ order term, can be combined to obtain a complete set of equations that yield the dynamical BCS approximation for the density response function $\chi_{nn}(\mathbf{q}, \omega)$. 
    All the details of the derivation can be found, for example, in \cite{PhysRevA.74.042717}. Here, we just mention the main result, which is crucial for implementing the methodology:
    \begin{equation} \label{eqn:density-density response function}
        \chi _ { n n }(\mathbf{q}, \omega) = - \frac { 1 } { V } \left\{ I ^ { \prime \prime } + \frac { \Delta ^ { 2 } } { I _ { 11 } I _ { 22 } - \omega ^ { 2 } I _ { 12 } } \left( 2 \omega ^ { 2 } I I _ { 12 } I ^ { \prime } - \omega ^ { 2 } I _ { 22 } I ^ { \prime 2 } - I ^ { 2 } I _ { 11 } \right) \right\}
    \end{equation}
    where $V$, the volume, is simply the number of lattice sites, and the various quantities that appear in the  response function are defined as follows:
    \begin{align}\label{eqn:I definitions}
       \begin{aligned} I ^ { \prime \prime } ( \mathbf { q } , \omega ) & := \sum _ { \mathbf { k } } \frac { E _ { + } + E _ { - } } { \left( E _ { + } + E _ { - } \right) ^ { 2 } - \omega ^ { 2 } } \left( \frac { E _ { - } E _ { + } - \xi - \xi _ { + } + \Delta } { E _ { - } E _ { + } } \right) \\ I ( \mathbf { q } , \omega ) & := \sum _ { \mathbf { k } } \frac { E _ { + } + E _ { - } } { \left( E _ { + } + E _ { - } \right) ^ { 2 } - \omega ^ { 2 } } \left( \frac { \xi _ { + } + \xi _ { - } } { E _ { - } E _ { + } } \right) \\ I ^ { \prime } ( \mathbf { q } , \omega ) & := \sum _ { \mathbf { k } } \frac { E _ { + } + E _ { - } } { \left( E _ { + } + E _ { - } \right) ^ { 2 } - \omega ^ { 2 } } \left( \frac { 1 } { E _ { - } E _ { + } } \right) \\
       I _ { 12 } ( \mathbf { q } , \omega ) & := \sum _ { \mathbf { k } } \frac { 1 } { \left( E _ { + } + E _ { - } \right) ^ { 2 } - \omega ^ { 2 } } \left( \frac { E + \xi_{-} + E_{-} \xi _ { + } } { E - E _ { + } } \right) \\ I _ { 11 } ( \mathbf { q } , \omega ) & := \sum _ { \mathbf { k } } \frac { E _ { + } + E _ { - } } { \left( E _ { + } + E _ { - } \right) ^ { 2 } - \omega ^ { 2 } } \left( \frac { E _ { - } E _ { + } + \xi _ { - } \xi _ { + } + \Delta ^ { 2 } } { E _ { - } E _ { + } } \right) - \frac { 1 } { E } \\ I _ { 22 } ( \mathbf { q } , \omega ) & := \sum _ { \mathbf { k } } \frac { E _ { + } + E _ { - } } { \left( E _ { + } + E _ { - } \right) ^ { 2 } - \omega ^ { 2 } } \left( \frac { E _ { - } E _ { + } + \xi _ { - } \xi _ { + } - \Delta ^ { 2 } } { E _ { - } E _ { + } } \right) - \frac { 1 } { E }\end{aligned}
    \end{align}
    where the shortcuts $E := E_k$, $E_{\pm} := E_{k \pm q/2}, \xi_{\pm}:=\xi_{k \pm q/2}$ where $\xi_k = e_k - \mu$
    have been used. Note that in the implementation, which we shall shortly discuss, for small lattices, care must be taken since the momentum $\mathbf{k} \pm \mathbf{q}/2$ may not be allowed by the periodic conditions imposed; fortunately, this issue is automatically "fixed" in very large lattices or at high momentum transfer. 
    
    Since we would like to compare with Quantum Monte Carlo results, we will focus on the density dynamic structure factor, $S(\mathbf{q}, \omega)$, which is a function of the momentum and energy transfer, and is related to the density response function via the celebrated fluctuation-dissipation theorem, 
    \begin{equation} \label{eqn:dynamic structure factor}
        S ( \mathbf { q } , \omega ) = \lim_{\eta \rightarrow 0^+ }\left( \frac{-1}{\pi n}\Im \left( \chi _ { n n } \left( \mathbf { q } , \omega + i \eta \right) \right) \right).
    \end{equation}
    Here, $\eta$ is a parameter introduced to avoid the poles that appear in eq. \ref{eqn:I definitions}. We will come back to this point below. 

%% file: dbcs_implementation.tex
    
    Before concluding our quick review of the methodology, we find useful to provide some practical suggestions to implement the dynamical BCS theory for a dilute 2D attractive Fermi gas; this theory yields a very valuable procedure to calculate dynamical correlation functions, in particular the density response function, which can be directly compared with experiment.
    The starting point is the choice of the interaction strength $\log(k_F a)$, which is related to the Hubbard interaction through the relation \eqref{eqn:HubbardU}.
    Implementing the actual dynamical BCS technique may be separated into two parts. First, the equilibrium chemical potential and pairing gap $(\Delta,\mu)$ must be obtained from an analysis of static case, solving the gap equation and the particle number equation. Second, the integrals appearing in the dynamical BCS approximation for the response function $\chi_{nn}(\mathbf{q}, \omega)$, which depend parametrically on $(\Delta,\mu)$, must be calculated numerically.
    Some care has to be taken due to the presence of many poles in the integrands.
    As it is clear from equation \ref{eqn:dynamic structure factor},
    the introduction of a real valued small parameter $\eta$ is necessary to avoid infinities through the substitution $\omega \to \omega + i \eta$. When studying finite lattices, a finite $\eta$ is required, where upon taking the continuum limit $\eta$ tends to zero. This means that $\eta$ does play role in calculating the response function for the two dimensional Fermi atomic gas. Below, we will propose a simple way to choose $\eta$, using the non-interacting limit.
    

%% file: calculations.tex
The dynamical BCS theory is a useful and computationally very cheap tool to compute the density response function and the dynamical structure factor for a Fermi gas. Naturally, it is an approximate scheme and an assessment is important. Auxiliary Field Quantum Monte Carlo simulations, on the other hand, are very robust, and in fact we can obtain exact numerical results \cite{PhysRevA.96.061601}; nevertheless, these approaches are more demanding and the exact predictions are limited to imaginary time domain, requiring delicate analytic continuation procedures to compare with experiments.
We choose a situation where a quantitative comparison can be made: we study 
a dilute system consisting of 18 particles moving on a square lattice with $25 \times 25$ lattice sites.

We focus on a few values of the interaction parameter $\log(k_Fa)$ that are meant to explore the physically interesting regimes, including the BEC-BCS crossover.
Namely, we scan the values $\log(k_Fa)=0.0$, $\log(k_Fa)=0.5$, $\log(k_Fa) = 1.0$, and $\log(k_Fa)=1.5$ \cite{PhysRevA.96.061601}, with the BEC regime being the former and BCS being the latter. Incidentally, we note that in cold atomic gas experiments, these parameters are easily controlled by the experimenter via Feshbach resonances.


Since the AFQMC predictions are only exact in imaginary time, the comparison between AFQMC and dynamical BCS predictions is fairest in imaginary time. This is done by examining dynamical BCS in imaginary time, which is obtained via the following Laplace transformation,
\begin{equation} \label{eqn:int scat function}
    F ( \mathbf { q } , \tau ) = \int _ { 0 } ^ {\infty } e ^ { - \tau \omega } S ( \mathbf { q } , \omega ) d \omega,
\end{equation} 
that can be performed with arbitrary precision. The intermediate scattering function, $F(\mathbf{q}, \tau)$, is a function of momentum transfer $\mathbf{q}$ and imaginary time, $\tau$. Comparing the results for $F(\mathbf{q}, \tau)$ from AFQMC and dynamical BCS
is a quantitative way to assess the validity of dynamical BCS and can be a very valuable asset to explore ways to improve dynamical BCS through renormalization of the parameters.

As mentioned above, in the framework of dynamical BCS, it is important to tune a parameter $\eta$, necessary to avoid infinities. This can introduce some arbitrariness in the approach. We find reasonable to choose $\eta$ in such a way that the non-interacting limit for $F ( \mathbf { q } , \tau )$ is correctly recovered as $\log(k_F a) \to + \infty$.

\begin{figure}[!hbt]
        \centering
        \subfloat[BEC regime $(\log(k_Fa)=0.0)$. ]{\label{fig:f1}{\includegraphics[width=0.5\textwidth]{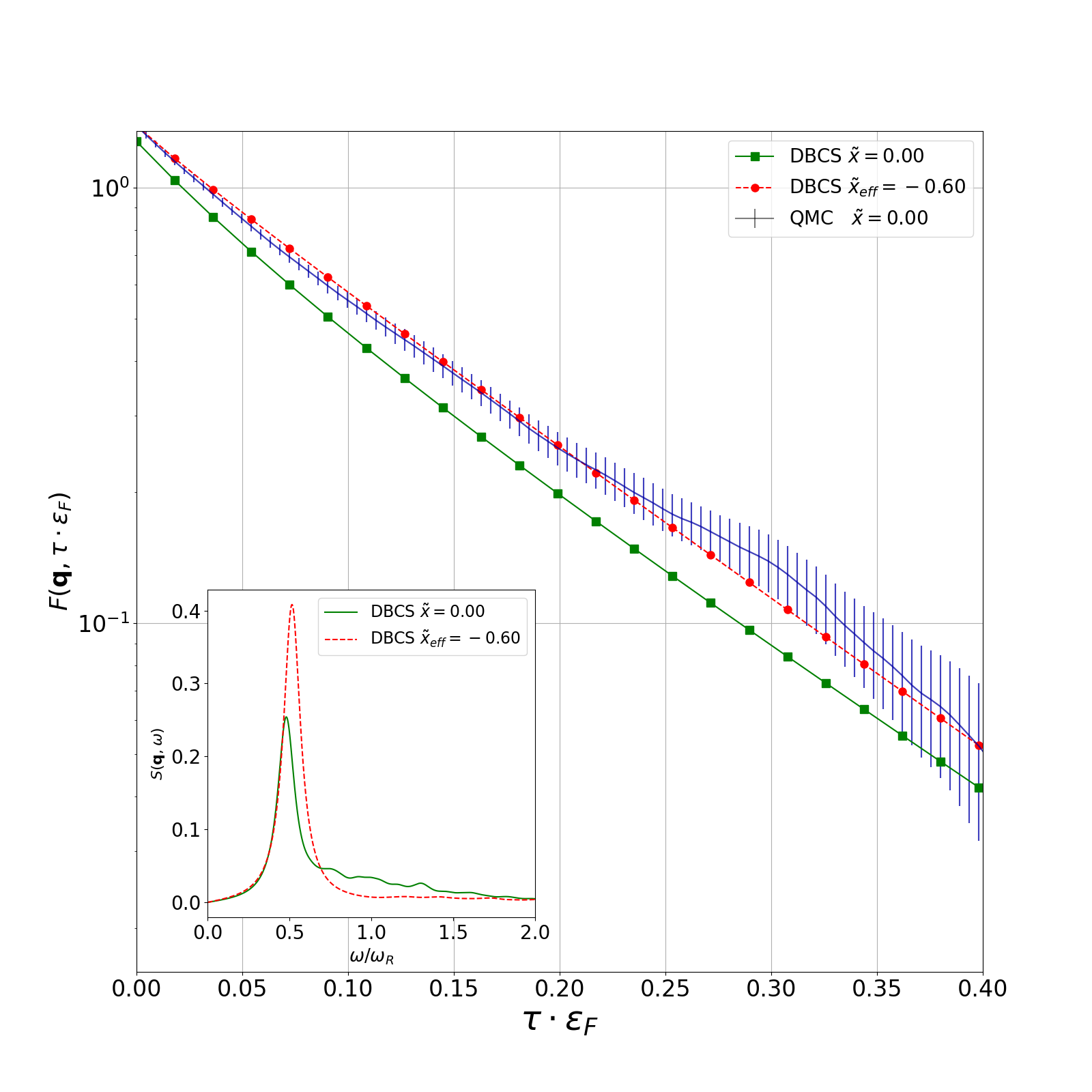}}}
        \subfloat[BEC-BCS crossover regime $(\log(k_Fa)=0.5)$]{\label{fig:f2}{\includegraphics[width=0.5\textwidth]{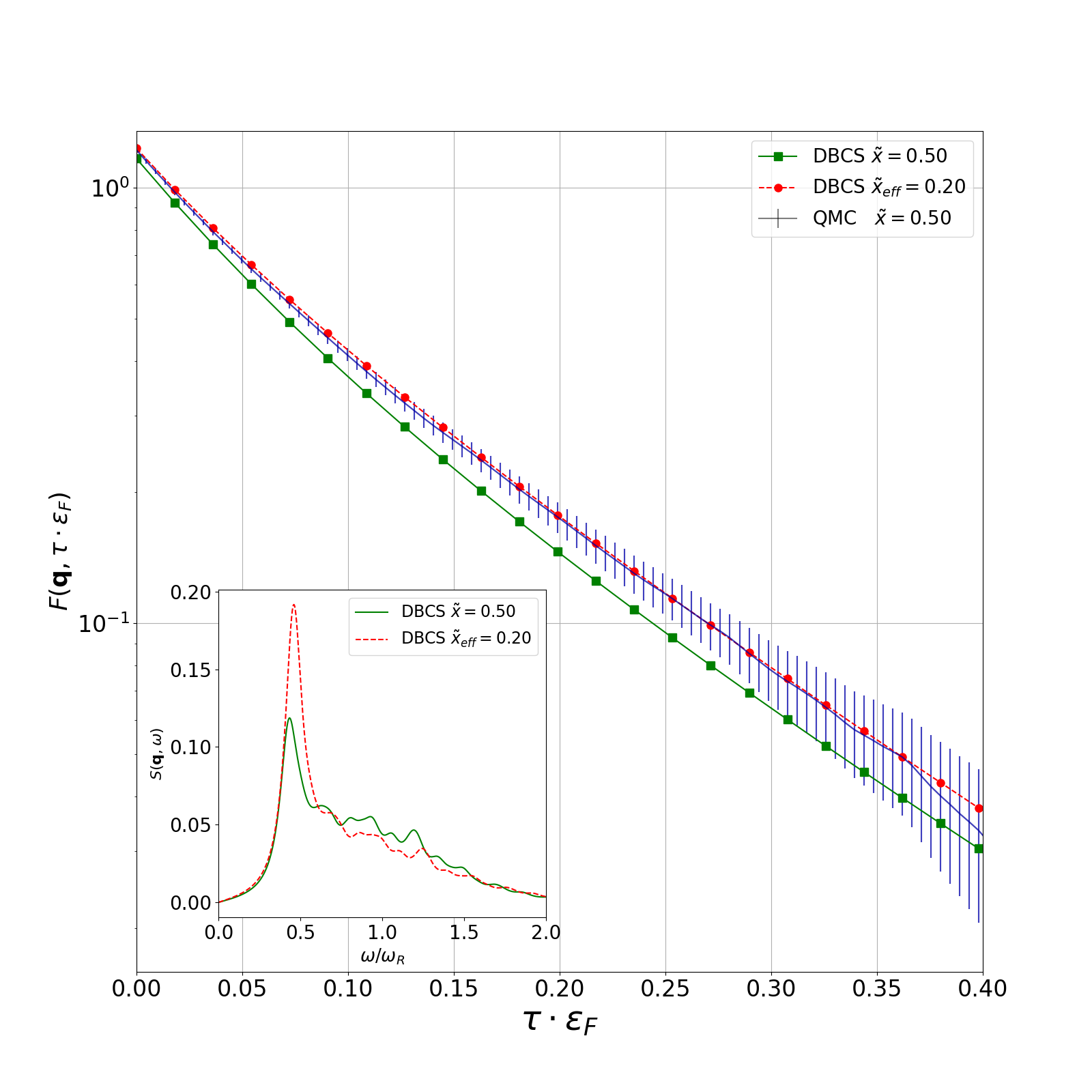}}}\hfill
        \subfloat[BEC-BCS crossover regime $(\log(k_Fa)=1.0)$]{\label{fig:f3}{\includegraphics[width=0.5\textwidth]{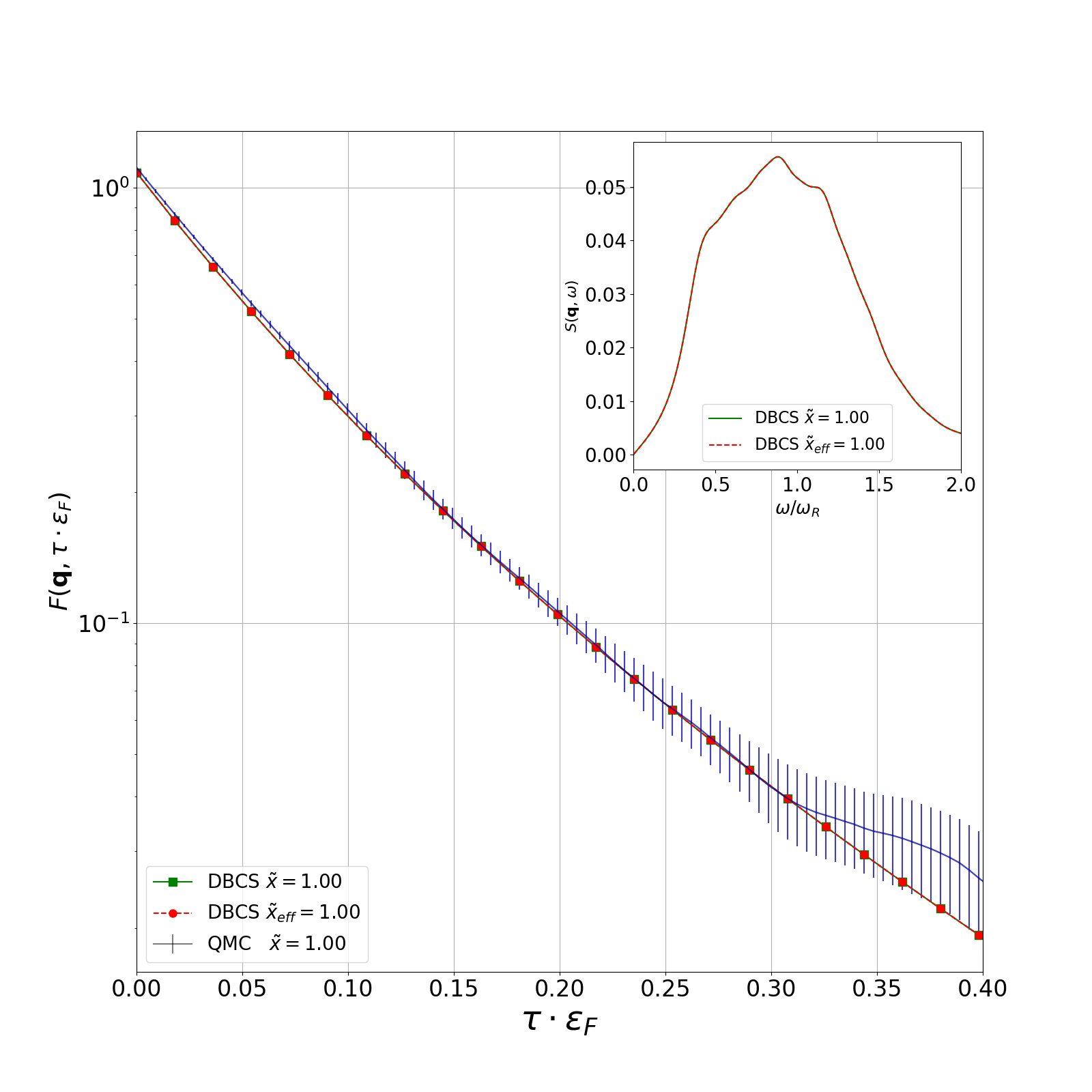}}}
        \subfloat[BCS regime $(\log(k_Fa)=1.5)$]{\label{fig:f4}{\includegraphics[width=0.5\textwidth]{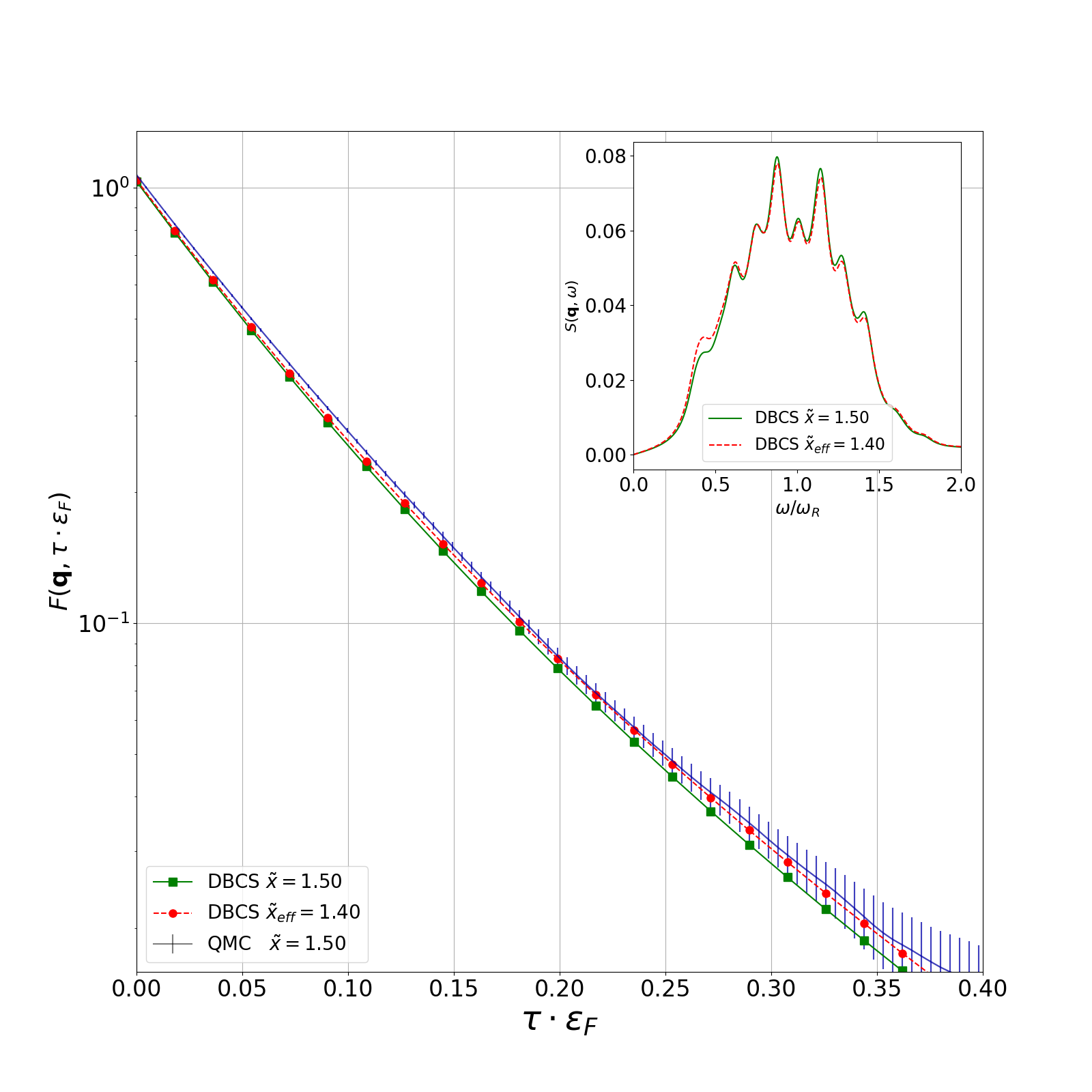}}}
        \caption{Comparison between dynamical BCS estimates and QMC results of the intermediate scattering function for a few values of the interaction strength $\tilde{x}=\log(k_Fa)$. The insets show the dynamical BCS predictions for the dynamic structure factor, $S(\mathbf{q}, \omega)$, in units of $\varepsilon_F^{-1}$. Calculations are done at $\eta=0.1$, choses to reproduce the non-interacting limit, and $|\mathbf{q}| = 4.0 k_F$. }
             \label{fig:renormalization_qmc}
    \end{figure}

   Building on a recent study \cite{responsefunctions}, we start our investigation at high momentum,
   $|\mathbf { q }| \simeq 4k_F$. This regime is important, since the dynamical structure factor is a very interesting probe to distinguish between the individual particles and the molecules. Indeed, at high momentum, the experimental probe sets in motion an individual particle or a molecule. In the insets of Fig.~\ref{fig:renormalization_qmc} we show the dynamical BCS estimations for $S(\mathbf { q }, \omega)$ as functions of $\omega/\omega_R(\mathbf { q })$, where $\omega_R(\mathbf { q }) = \frac{\hbar^2 |\mathbf { q }|^2}{2 m}$ is the recoil energy of a free atom. As discussed in \cite{PhysRevA.96.061601}, in the BEC regime the peak is located at $0.5 \, \omega_R(\mathbf { q })$, corresponding to a mass equal to $2m$, which is exactly the mass of a tightly bound molecule. On the other hand, in the BCS regime, the peak is located at $\omega_R(\mathbf { q })$, as an almost unbound individual atom responds to the perturbation.
   The focus of this paper is on the comparison with AFQMC, and so the main panels of Fig.~\ref{fig:renormalization_qmc} show the intermediate scattering function for the different regimes. Each panel displays the unbiased AFQMC result, together with the statistical uncertainties, and two dynamical BCS predictions: one of them (the solid green line with square markers) is obtained implementing the procedure exactly as described in this paper. The other, dotted red line with circle markers, on the other hand, is obtained by using an effective renormalized interaction parameter, say $\tilde{x}_{eff}$, which is allowed to be different from the physical $\log(k_F a)$. We choose $\tilde{x}_{eff}$ as the value that minimizes the discrepancy between the dynamical BCS prediction and the AFQMC exact result. We see that, for all the considered values of the physical $\log(k_F a)$, we are able to choose $\tilde{x}_{eff}$ such that intermediate scattering function is in reasonable agreement with the AFQMC result within the statistical errors, at least for $\tau \varepsilon_F \geq 0.15$.
   We stress that the large values of $\tau \varepsilon_F$ are more significant to describe the low energy dynamics, and so it is more desirable to have agreement at large $\tau \varepsilon_F$; the short imaginary time behavior is determined by several high energy excited states and can be very hard to predict.
   We observe that, for all the considered values of $\log(k_F a)$, the optimal renormalized interaction parameter turns out to be always smaller than or equal to the physical one. The optimal values are presented in the figure.  This means that, in order to correct the dynamical BCS theory, we need to consider an effective interaction that is stronger than the physical one, at least for the considered regimes and for high value of the momentum.
   This seems to suggest that, if we probe the system at high momentum, the physical correlations somehow make the pairing stronger.

    Although the high momentum behavior is very interesting, as it is a direct probe of the crossover from a gas of Bose molecules to a superfluid of Fermi atoms, it is intriguing to explore if the renormalization strategy can be applied also for lower values of the momentum. We expect that the situation becomes immediately more complicated, due to the particle-hole excitations that play a major role for intermediate values of $|\mathbf{q}|$.
    We discovered that, in the BEC and in the crossover regime, there is no way to reconcile the AFQMC result with the dynamical BCS predictions: renormalizing a single parameter is not enough to correct the predictions. More advanced approaches need to be devised and they can be interesting topics for future research.
    On the other hand, in the BCS regime, for $\log(k_F a) \geq 1.5$, it is still possible to renormalize the dynamical BCS theory to get an effective interaction parameter $\tilde{x}_{eff}$, which appears to depend on $|\mathbf{q}|$.

    
    \begin{figure}[!hbt]
        \centering
        \subfloat[Weakly  interacting regime $(\log(k_Fa)=1.5)$]{\label{fig:f5}{\includegraphics[width=0.5\textwidth]{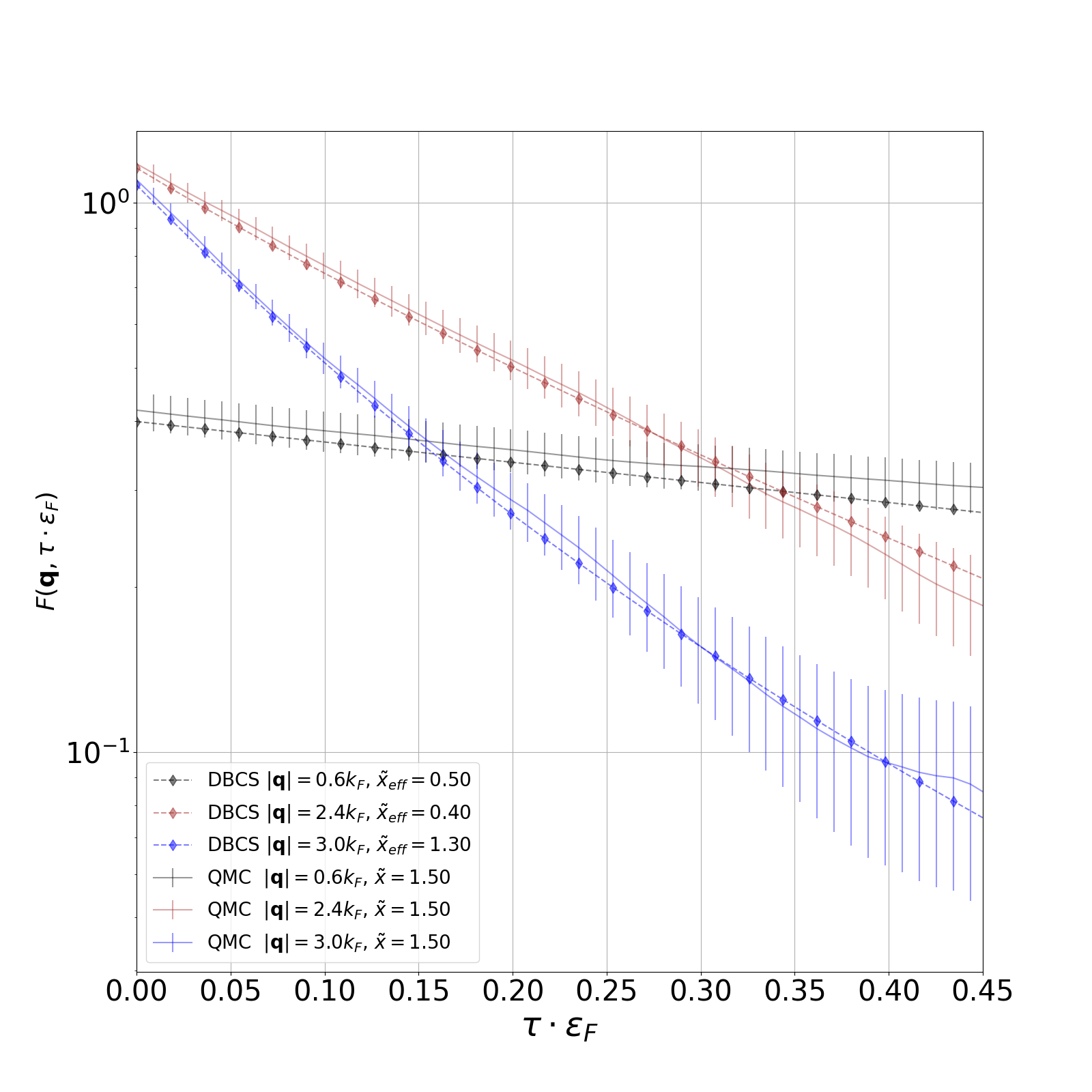}}}
        \subfloat[Weakly interacting regime $(\log(k_Fa)=3.5)$]{\label{fig:f6}{\includegraphics[width=0.5\textwidth]{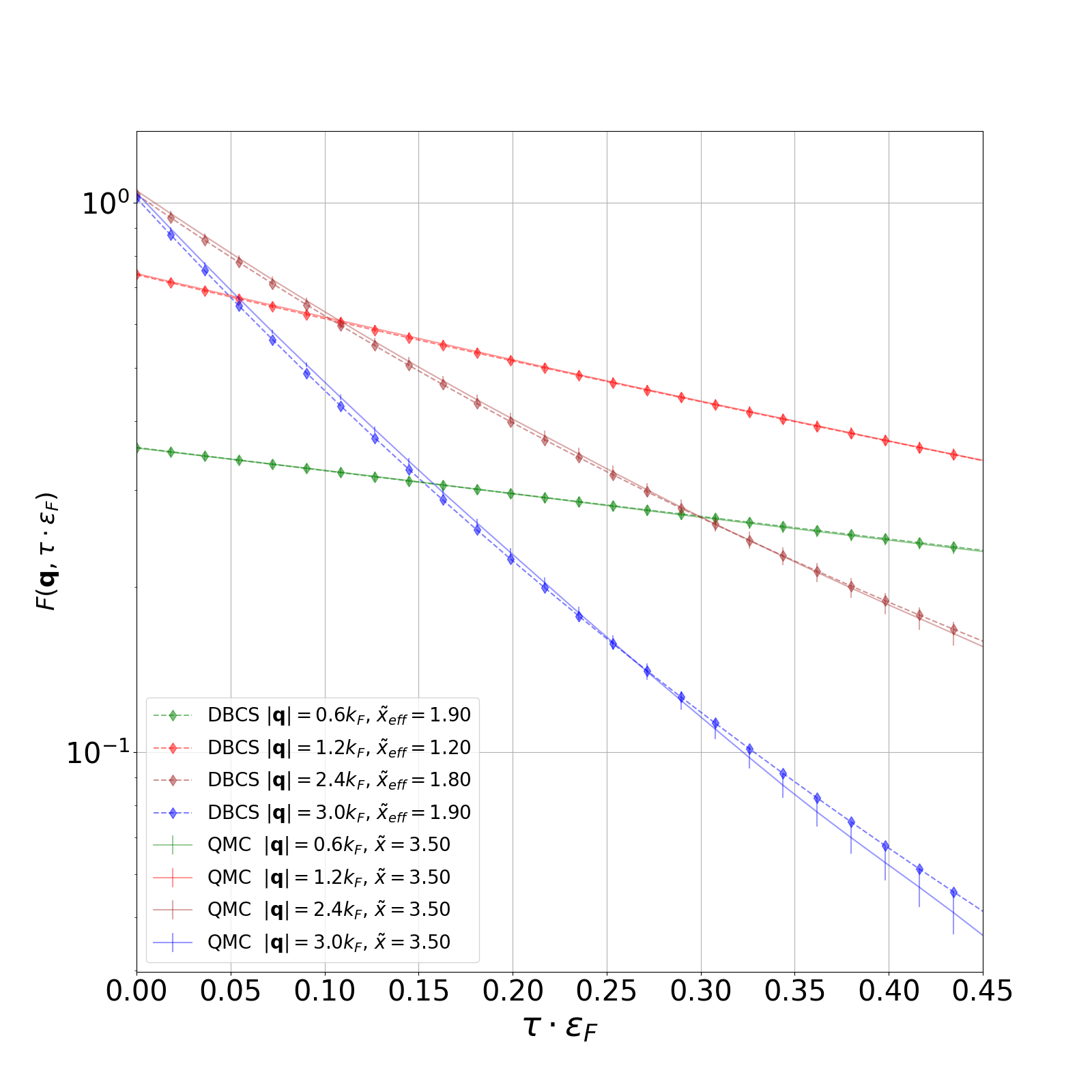}}}\hfill
        \caption{Comparison between AFQMC results for the intermediate scattering functions, shown with error bars, and dynamical BCS  predictions in two interaction regimes ($\tilde{x}=\log(k_Fa)=1.5, 3.5$) for a few values of the momentum transfer.   }
          \label{fig:bcs_regime_allq}
    \end{figure}
    The results are shown in Fig.~\ref{fig:bcs_regime_allq}, where we consider $\log(k_Fa) = 1.5$ and $\log(k_F a) = 3.5$. The agreement is good, although we cannot extract a significant trend for the dependence of $\tilde{x}_{eff}$ on the momentum: it appears that, empirically, we need to find the optimal value for each momentum.
    
    The results we obtain are encouraging and show that, in several situations, it is possible to implement a dynamical BCS scheme with a renormalized parameter whose prediction for the intermediate scattering function in good agreement with the exact result. We stress that this is a simple first step toward a systematic approach to the formidable task of calculating  the density response function of an interacting quantum many-body system. Nevertheless, we are confident that the interface between simple approaches and sophisticated many-body approaches like AFQMC can be the method of choice to study many-body dynamics. 

%% file: conclusion.tex
Focusing on a 2D dilute attractive Fermi gas, we have used unbiased Quantum Monte Carlo calculations for the intermediate scattering function in imaginary time to assess the accuracy of the dynamical BCS predictions. Moreover, we explored the possibility to renormalize the interactions in the dynamical BCS theory to improve the agreement with exact results. We found that, at high momentum, it is possible to choose an effective parameter that makes the agreement very good, at least for large values of the imaginary time. This is true both in the BEC and BCS regimes, and even in the strongly correlated crossover regime.
On the other hand, at smaller momenta, we found empirically that it is possible to recover good agreement by choosing a momentum dependent effective parameter only in the BCS regime, while for $\log(k_F a) < 1$ it appears that the renormalization of a single parameter does not significantly improve the accuracy of the predictions.

Due to the well known importance of response functions, we are confident that our results provide a significant first step in the direction of using exact results from sophisticated correlated methodologies, like Quantum Monte Carlo, as an asset to assess and improve simpler theories that may allow us to study dynamical properties of bulk systems in real time. 
We hope that our research will stimulate other groups to investigate these kind of techniques in a systematic way.

Quantum Monte Carlo computing was carried out at the Extreme Science and Engineering Discovery Environment (XSEDE), which is supported by National Science Foundation grant number ACI-1053575. One of us, E. V., would like to aknowledge useful discussions with Shiwei Zhang.

%% file: main.bbl
\begin{thebibliography}{16}%
\makeatletter
\providecommand \@ifxundefined [1]{%
 \@ifx{#1\undefined}
}%
\providecommand \@ifnum [1]{%
 \ifnum #1\expandafter \@firstoftwo
 \else \expandafter \@secondoftwo
 \fi
}%
\providecommand \@ifx [1]{%
 \ifx #1\expandafter \@firstoftwo
 \else \expandafter \@secondoftwo
 \fi
}%
\providecommand \natexlab [1]{#1}%
\providecommand \enquote  [1]{``#1''}%
\providecommand \bibnamefont  [1]{#1}%
\providecommand \bibfnamefont [1]{#1}%
\providecommand \citenamefont [1]{#1}%
\providecommand \href@noop [0]{\@secondoftwo}%
\providecommand \href [0]{\begingroup \@sanitize@url \@href}%
\providecommand \@href[1]{\@@startlink{#1}\@@href}%
\providecommand \@@href[1]{\endgroup#1\@@endlink}%
\providecommand \@sanitize@url [0]{\catcode `\\12\catcode `\$12\catcode
  `\&12\catcode `\#12\catcode `\^12\catcode `\_12\catcode `\%12\relax}%
\providecommand \@@startlink[1]{}%
\providecommand \@@endlink[0]{}%
\providecommand \url  [0]{\begingroup\@sanitize@url \@url }%
\providecommand \@url [1]{\endgroup\@href {#1}{\urlprefix }}%
\providecommand \urlprefix  [0]{URL }%
\providecommand \Eprint [0]{\href }%
\providecommand \doibase [0]{http://dx.doi.org/}%
\providecommand \selectlanguage [0]{\@gobble}%
\providecommand \bibinfo  [0]{\@secondoftwo}%
\providecommand \bibfield  [0]{\@secondoftwo}%
\providecommand \translation [1]{[#1]}%
\providecommand \BibitemOpen [0]{}%
\providecommand \bibitemStop [0]{}%
\providecommand \bibitemNoStop [0]{.\EOS\space}%
\providecommand \EOS [0]{\spacefactor3000\relax}%
\providecommand \BibitemShut  [1]{\csname bibitem#1\endcsname}%
\let\auto@bib@innerbib\@empty
\bibitem [{\citenamefont {Giorgini}\ \emph {et~al.}(2008)\citenamefont
  {Giorgini}, \citenamefont {Pitaevskii},\ and\ \citenamefont
  {Stringari}}]{RevModPhys.80.1215}%
  \BibitemOpen
  \bibfield  {author} {\bibinfo {author} {\bibfnamefont {S.}~\bibnamefont
  {Giorgini}}, \bibinfo {author} {\bibfnamefont {L.~P.}\ \bibnamefont
  {Pitaevskii}}, \ and\ \bibinfo {author} {\bibfnamefont {S.}~\bibnamefont
  {Stringari}},\ }\href {\doibase 10.1103/RevModPhys.80.1215} {\bibfield
  {journal} {\bibinfo  {journal} {Rev. Mod. Phys.}\ }\textbf {\bibinfo {volume}
  {80}},\ \bibinfo {pages} {1215} (\bibinfo {year} {2008})}\BibitemShut
  {NoStop}%
\bibitem [{\citenamefont {Bloch}\ \emph {et~al.}(2008)\citenamefont {Bloch},
  \citenamefont {Dalibard},\ and\ \citenamefont {Zwerger}}]{RevModPhys.80.885}%
  \BibitemOpen
  \bibfield  {author} {\bibinfo {author} {\bibfnamefont {I.}~\bibnamefont
  {Bloch}}, \bibinfo {author} {\bibfnamefont {J.}~\bibnamefont {Dalibard}}, \
  and\ \bibinfo {author} {\bibfnamefont {W.}~\bibnamefont {Zwerger}},\ }\href
  {\doibase 10.1103/RevModPhys.80.885} {\bibfield  {journal} {\bibinfo
  {journal} {Rev. Mod. Phys.}\ }\textbf {\bibinfo {volume} {80}},\ \bibinfo
  {pages} {885} (\bibinfo {year} {2008})}\BibitemShut {NoStop}%
\bibitem [{\citenamefont {Braaten}\ \emph {et~al.}(2008)\citenamefont
  {Braaten}, \citenamefont {Kang},\ and\ \citenamefont
  {Platter}}]{PhysRevA.78.053606}%
  \BibitemOpen
  \bibfield  {author} {\bibinfo {author} {\bibfnamefont {E.}~\bibnamefont
  {Braaten}}, \bibinfo {author} {\bibfnamefont {D.}~\bibnamefont {Kang}}, \
  and\ \bibinfo {author} {\bibfnamefont {L.}~\bibnamefont {Platter}},\ }\href
  {\doibase 10.1103/PhysRevA.78.053606} {\bibfield  {journal} {\bibinfo
  {journal} {Phys. Rev. A}\ }\textbf {\bibinfo {volume} {78}},\ \bibinfo
  {pages} {053606} (\bibinfo {year} {2008})}\BibitemShut {NoStop}%
\bibitem [{\citenamefont {Lee}\ \emph {et~al.}(2006)\citenamefont {Lee},
  \citenamefont {Nagaosa},\ and\ \citenamefont {Wen}}]{RevModPhys.78.17}%
  \BibitemOpen
  \bibfield  {author} {\bibinfo {author} {\bibfnamefont {P.~A.}\ \bibnamefont
  {Lee}}, \bibinfo {author} {\bibfnamefont {N.}~\bibnamefont {Nagaosa}}, \ and\
  \bibinfo {author} {\bibfnamefont {X.-G.}\ \bibnamefont {Wen}},\ }\href
  {\doibase 10.1103/RevModPhys.78.17} {\bibfield  {journal} {\bibinfo
  {journal} {Rev. Mod. Phys.}\ }\textbf {\bibinfo {volume} {78}},\ \bibinfo
  {pages} {17} (\bibinfo {year} {2006})}\BibitemShut {NoStop}%
\bibitem [{\citenamefont {Qi}\ and\ \citenamefont
  {Zhang}(2011)}]{RevModPhys.83.1057}%
  \BibitemOpen
  \bibfield  {author} {\bibinfo {author} {\bibfnamefont {X.-L.}\ \bibnamefont
  {Qi}}\ and\ \bibinfo {author} {\bibfnamefont {S.-C.}\ \bibnamefont {Zhang}},\
  }\href {\doibase 10.1103/RevModPhys.83.1057} {\bibfield  {journal} {\bibinfo
  {journal} {Rev. Mod. Phys.}\ }\textbf {\bibinfo {volume} {83}},\ \bibinfo
  {pages} {1057} (\bibinfo {year} {2011})}\BibitemShut {NoStop}%
\bibitem [{\citenamefont {Yu}\ and\ \citenamefont
  {Bulgac}(2003)}]{PhysRevLett.90.161101}%
  \BibitemOpen
  \bibfield  {author} {\bibinfo {author} {\bibfnamefont {Y.}~\bibnamefont
  {Yu}}\ and\ \bibinfo {author} {\bibfnamefont {A.}~\bibnamefont {Bulgac}},\
  }\href {\doibase 10.1103/PhysRevLett.90.161101} {\bibfield  {journal}
  {\bibinfo  {journal} {Phys. Rev. Lett.}\ }\textbf {\bibinfo {volume} {90}},\
  \bibinfo {pages} {161101} (\bibinfo {year} {2003})}\BibitemShut {NoStop}%
\bibitem [{\citenamefont {Martiyanov}\ \emph {et~al.}(2010)\citenamefont
  {Martiyanov}, \citenamefont {Makhalov},\ and\ \citenamefont
  {Turlapov}}]{PhysRevLett.105.030404}%
  \BibitemOpen
  \bibfield  {author} {\bibinfo {author} {\bibfnamefont {K.}~\bibnamefont
  {Martiyanov}}, \bibinfo {author} {\bibfnamefont {V.}~\bibnamefont
  {Makhalov}}, \ and\ \bibinfo {author} {\bibfnamefont {A.}~\bibnamefont
  {Turlapov}},\ }\href {\doibase 10.1103/PhysRevLett.105.030404} {\bibfield
  {journal} {\bibinfo  {journal} {Phys. Rev. Lett.}\ }\textbf {\bibinfo
  {volume} {105}},\ \bibinfo {pages} {030404} (\bibinfo {year}
  {2010})}\BibitemShut {NoStop}%
\bibitem [{\citenamefont {Galea}\ \emph {et~al.}(2017)\citenamefont {Galea},
  \citenamefont {Zielinski}, \citenamefont {Gandolfi},\ and\ \citenamefont
  {Gezerlis}}]{Galea2017}%
  \BibitemOpen
  \bibfield  {author} {\bibinfo {author} {\bibfnamefont {A.}~\bibnamefont
  {Galea}}, \bibinfo {author} {\bibfnamefont {T.}~\bibnamefont {Zielinski}},
  \bibinfo {author} {\bibfnamefont {S.}~\bibnamefont {Gandolfi}}, \ and\
  \bibinfo {author} {\bibfnamefont {A.}~\bibnamefont {Gezerlis}},\ }\href
  {\doibase 10.1007/s10909-017-1803-1} {\bibfield  {journal} {\bibinfo
  {journal} {Journal of Low Temperature Physics}\ }\textbf {\bibinfo {volume}
  {189}},\ \bibinfo {pages} {451} (\bibinfo {year} {2017})}\BibitemShut
  {NoStop}%
\bibitem [{\citenamefont {Hoinka}\ \emph {et~al.}(2012)\citenamefont {Hoinka},
  \citenamefont {Lingham}, \citenamefont {Delehaye},\ and\ \citenamefont
  {Vale}}]{PhysRevLett.109.050403}%
  \BibitemOpen
  \bibfield  {author} {\bibinfo {author} {\bibfnamefont {S.}~\bibnamefont
  {Hoinka}}, \bibinfo {author} {\bibfnamefont {M.}~\bibnamefont {Lingham}},
  \bibinfo {author} {\bibfnamefont {M.}~\bibnamefont {Delehaye}}, \ and\
  \bibinfo {author} {\bibfnamefont {C.~J.}\ \bibnamefont {Vale}},\ }\href
  {\doibase 10.1103/PhysRevLett.109.050403} {\bibfield  {journal} {\bibinfo
  {journal} {Phys. Rev. Lett.}\ }\textbf {\bibinfo {volume} {109}},\ \bibinfo
  {pages} {050403} (\bibinfo {year} {2012})}\BibitemShut {NoStop}%
\bibitem [{\citenamefont {Shi}\ \emph {et~al.}(2015)\citenamefont {Shi},
  \citenamefont {Chiesa},\ and\ \citenamefont {Zhang}}]{PhysRevA.92.033603}%
  \BibitemOpen
  \bibfield  {author} {\bibinfo {author} {\bibfnamefont {H.}~\bibnamefont
  {Shi}}, \bibinfo {author} {\bibfnamefont {S.}~\bibnamefont {Chiesa}}, \ and\
  \bibinfo {author} {\bibfnamefont {S.}~\bibnamefont {Zhang}},\ }\href
  {\doibase 10.1103/PhysRevA.92.033603} {\bibfield  {journal} {\bibinfo
  {journal} {Phys. Rev. A}\ }\textbf {\bibinfo {volume} {92}},\ \bibinfo
  {pages} {033603} (\bibinfo {year} {2015})}\BibitemShut {NoStop}%
\bibitem [{\citenamefont {Vitali}\ \emph
  {et~al.}(2017{\natexlab{a}})\citenamefont {Vitali}, \citenamefont {Shi},
  \citenamefont {Qin},\ and\ \citenamefont {Zhang}}]{PhysRevA.96.061601}%
  \BibitemOpen
  \bibfield  {author} {\bibinfo {author} {\bibfnamefont {E.}~\bibnamefont
  {Vitali}}, \bibinfo {author} {\bibfnamefont {H.}~\bibnamefont {Shi}},
  \bibinfo {author} {\bibfnamefont {M.}~\bibnamefont {Qin}}, \ and\ \bibinfo
  {author} {\bibfnamefont {S.}~\bibnamefont {Zhang}},\ }\href {\doibase
  10.1103/PhysRevA.96.061601} {\bibfield  {journal} {\bibinfo  {journal} {Phys.
  Rev. A}\ }\textbf {\bibinfo {volume} {96}},\ \bibinfo {pages} {061601}
  (\bibinfo {year} {2017}{\natexlab{a}})}\BibitemShut {NoStop}%
\bibitem [{\citenamefont {Bertaina}\ \emph {et~al.}(2017)\citenamefont
  {Bertaina}, \citenamefont {Galli},\ and\ \citenamefont
  {Vitali}}]{statisticalcomp}%
  \BibitemOpen
  \bibfield  {author} {\bibinfo {author} {\bibfnamefont {G.}~\bibnamefont
  {Bertaina}}, \bibinfo {author} {\bibfnamefont {D.}~\bibnamefont {Galli}}, \
  and\ \bibinfo {author} {\bibfnamefont {E.}~\bibnamefont {Vitali}},\ }\href
  {\doibase 10.1080/23746149.2017.1288585} {\bibfield  {journal} {\bibinfo
  {journal} {Advances in Physics: X}\ }\textbf {\bibinfo {volume} {2}},\
  \bibinfo {pages} {302} (\bibinfo {year} {2017})}\BibitemShut {NoStop}%
\bibitem [{\citenamefont {Vitali}\ \emph {et~al.}(2016)\citenamefont {Vitali},
  \citenamefont {Shi}, \citenamefont {Qin},\ and\ \citenamefont
  {Zhang}}]{PhysRevB.94.085140}%
  \BibitemOpen
  \bibfield  {author} {\bibinfo {author} {\bibfnamefont {E.}~\bibnamefont
  {Vitali}}, \bibinfo {author} {\bibfnamefont {H.}~\bibnamefont {Shi}},
  \bibinfo {author} {\bibfnamefont {M.}~\bibnamefont {Qin}}, \ and\ \bibinfo
  {author} {\bibfnamefont {S.}~\bibnamefont {Zhang}},\ }\href {\doibase
  10.1103/PhysRevB.94.085140} {\bibfield  {journal} {\bibinfo  {journal} {Phys.
  Rev. B}\ }\textbf {\bibinfo {volume} {94}},\ \bibinfo {pages} {085140}
  (\bibinfo {year} {2016})}\BibitemShut {NoStop}%
\bibitem [{\citenamefont {Vitali}\ \emph
  {et~al.}(2017{\natexlab{b}})\citenamefont {Vitali}, \citenamefont {Shi},
  \citenamefont {Qin},\ and\ \citenamefont {Zhang}}]{responsefunctions}%
  \BibitemOpen
  \bibfield  {author} {\bibinfo {author} {\bibfnamefont {E.}~\bibnamefont
  {Vitali}}, \bibinfo {author} {\bibfnamefont {H.}~\bibnamefont {Shi}},
  \bibinfo {author} {\bibfnamefont {M.}~\bibnamefont {Qin}}, \ and\ \bibinfo
  {author} {\bibfnamefont {S.}~\bibnamefont {Zhang}},\ }\href {\doibase
  10.1007/s10909-017-1805-z} {\bibfield  {journal} {\bibinfo  {journal}
  {Journal of Low Temperature Physics}\ } (\bibinfo {year}
  {2017}{\natexlab{b}}),\ 10.1007/s10909-017-1805-z}\BibitemShut {NoStop}%
\bibitem [{\citenamefont {Combescot}\ \emph {et~al.}(2006)\citenamefont
  {Combescot}, \citenamefont {Kagan},\ and\ \citenamefont
  {Stringari}}]{PhysRevA.74.042717}%
  \BibitemOpen
  \bibfield  {author} {\bibinfo {author} {\bibfnamefont {R.}~\bibnamefont
  {Combescot}}, \bibinfo {author} {\bibfnamefont {M.~Y.}\ \bibnamefont
  {Kagan}}, \ and\ \bibinfo {author} {\bibfnamefont {S.}~\bibnamefont
  {Stringari}},\ }\href {\doibase 10.1103/PhysRevA.74.042717} {\bibfield
  {journal} {\bibinfo  {journal} {Phys. Rev. A}\ }\textbf {\bibinfo {volume}
  {74}},\ \bibinfo {pages} {042717} (\bibinfo {year} {2006})}\BibitemShut
  {NoStop}%
\bibitem [{\citenamefont {Werner}\ and\ \citenamefont
  {Castin}(2012)}]{PhysRevA.86.013626}%
  \BibitemOpen
  \bibfield  {author} {\bibinfo {author} {\bibfnamefont {F.}~\bibnamefont
  {Werner}}\ and\ \bibinfo {author} {\bibfnamefont {Y.}~\bibnamefont
  {Castin}},\ }\href {\doibase 10.1103/PhysRevA.86.013626} {\bibfield
  {journal} {\bibinfo  {journal} {Phys. Rev. A}\ }\textbf {\bibinfo {volume}
  {86}},\ \bibinfo {pages} {013626} (\bibinfo {year} {2012})}\BibitemShut
  {NoStop}%
\end{thebibliography}
